\documentclass{ws-procs9x6}

\begin{document}

\title{THE 2PPI EXPANSION: DYNAMICAL MASS GENERATION AND VACUUM ENERGY\footnote{\uppercase{T}alk given
 by \uppercase{D}. \uppercase{D}udal at the \uppercase{I}nternational
\uppercase{C}onference on \uppercase{C}olor
\uppercase{C}onfinement and \uppercase{H}adrons in
\uppercase{Q}uantum \uppercase{C}hromodynamics,
\uppercase{C}onfinement 2003, \uppercase{TIT}ech \&
\uppercase{R}iken, \uppercase{T}okyo, \uppercase{J}apan,
\uppercase{J}uly 21-24, 2003.}}

\author{D. Dudal\footnote{\uppercase{R}esearch \uppercase{A}ssistant of the
 \uppercase{F}und for \uppercase{S}cientific \uppercase{R}esearch-\uppercase{F}landers, \uppercase{B}elgium.} and H. VERSCHELDE}

\address{Ghent University\\
Department of Mathematical Physics and Astronomy\\
Krijgslaan 281-S9, B-9000 Gent, Belgium\\
E-mail: david.dudal@ugent.be,henri.verschelde@ugent.be}

\author{R.~E. Browne and J.~A. Gracey}

\address{Theoretical Physics Division, Department of Mathematical Sciences\\
University of Liverpool\\
P.O. Box 147, Liverpool, L69 3BX, United Kingdom\\
E-mail: rebrowne@amtp.liv.ac.uk, jag@amtp.liv.ac.uk}


\maketitle \vspace{-13cm} \hfill LTH--601 \vspace{13cm}

\abstracts{We discuss the $2PPI$ expansion, a summation of the
bubble graphs up to all orders, by means of the $2D$ Gross-Neveu
toy model, whose exact mass gap and vacuum energy are known. Then
we use the expansion to give analytical evidence that a dimension
two gluon condensate exists for pure Yang-Mills in the Landau
gauge. This $\left\langle A_\mu^a A_\mu^a\right\rangle$ condensate
consequently gives rise to a dynamical gluon mass.}

\section{Introduction}
Lately, there was growing evidence for the existence of a
condensate of mass dimension two in Yang-Mills (YM) theories in
the Landau gauge. An obvious candidate for such a condensate is
$\left\langle A_{\mu}^a A_{\mu}^a\right\rangle$. The
phenomenological background of this type of condensate can be
found in \cite{Chetyrkin:1998yr,Gubarev:2000eu,Gubarev:2000nz}.
Also lattice simulations indicated a non-zero condensate
$\left\langle A_{\mu}^a A_{\mu}^a\right\rangle$
\cite{Boucaud:2001st}. See \cite{Dudal:2003tc} for an overview of
recent results.\\\\Thinking of simpler models like massless
$\lambda \phi^{4}$ or Gross-Neveu \cite{Gross:jv} and the role
played by their quartic interaction in the formation of a (local)
composite (in particular, containing two fields) condensate and
the consequent dynamical mass generation for the originally
massless fields \cite{Gross:jv,Dudal:2002zn}, it is clear the
possibility exists that the quartic gluon interaction gives rise
to a two field composite operator
condensate in YM (QCD) and mass generation for the gluons too.\\\\
In section 2, we give a short review of the $2PPI$ expansion by
means of the $2D$ Gross-Neveu model. In section 3, we present our
results for $\left\langle A^2\right\rangle$ for $SU(N)$ Yang-Mills
theory in the Landau gauge

\section{The Gross-Neveu model}
The $U(N)$ invariant Gross-Neveu Lagrangian in $2D$ Euclidean
space time reads
\begin{equation}\label{1}
    \mathcal{L}=\overline{\psi}\partial\hspace{-0.2cm}/\psi-\frac{1}{2}g^{2}\left(\overline{\psi}\psi\right)^{2}.
\end{equation}
This model possesses a discrete chiral symmetry
$\psi\rightarrow\gamma_{5}\psi$, imposing
$\left\langle\overline{\psi}\psi\right\rangle=0$ perturbatively.
We focus on the topology of vacuum diagrams. We can divide them
into 2 disjoint classes: those diagrams falling apart in 2
separate pieces when 2 lines meeting at the same point are cut. We
call those \textit{2-point-particle-reducible} or $2PPR$. The
second diagram of Figure 1 is a $2PPR$ diagram. The other type is
the complement of the $2PPR$ class, we call these diagrams
\textit{2-point-particle-irreducible} ($2PPI$) diagrams. The first
and third diagram of Figure 1 are $2PPI$ ones.
\begin{figure}[ht]
\epsfxsize=4cm   
\centerline{\epsfxsize=3.9in\epsfbox{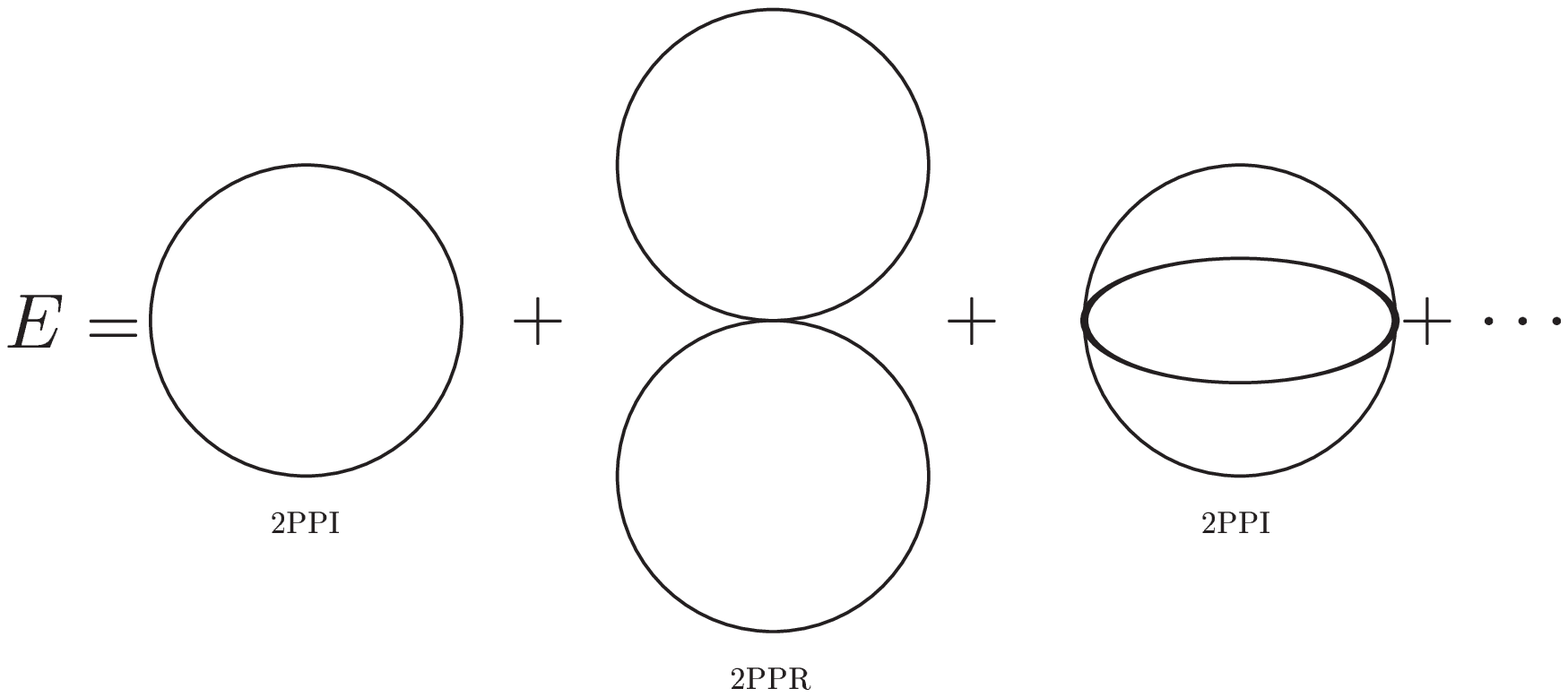}} \caption{The
vacuum energy as a sum of $2PPI$ and $2PPR$ diagrams.
\label{inter}}
\end{figure}
We could now remove all $2PPR$ bubbles from the diagrammatic sum
building up the vacuum energy by summing them in an effective mass
$m$. Defining $\Delta = \langle\overline{\psi}\psi\rangle$, it can
be shown\cite{Dudal:2002zn} that
\begin{equation}\label{3}
    m=-g^{2}\left(1-\frac{1}{2N}\right)\Delta.
\end{equation}
Then the $2PPI$ vacuum energy $E_{2PPI}$ is given by the sum of
all $2PPI$ vacuum diagrams, now with a mass $m$ running in the
loops. It is important to notice that $E_{2PPI}$ is \emph{not} the
vacuum energy due to a double counting ambiguity, already visible
in the second diagram of Figure 1: each diagram can be seen as an
insertion on the other one. This can be solved by considering
$\frac{dE}{dg^{2}}$ instead of $E$. The $g^{2}$ derivative can hit
$2PPR$ or $2PPI$ vertex.
\begin{figure}[ht]
\epsfxsize=4cm   
\centerline{\epsfxsize=3.9in\epsfbox{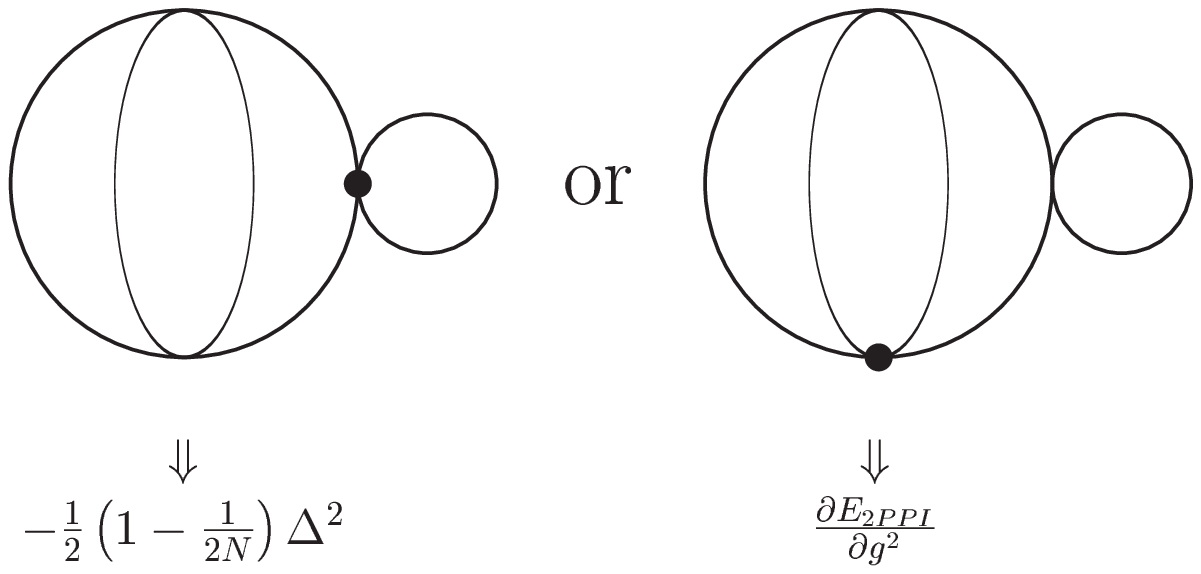}} \caption{The
$2PPI$ vacuum energy $E_{2PPI}$. \label{inter}}
\end{figure}
Hence
\begin{equation}\label{4}
    \frac{dE}{dg^{2}}=-\frac{1}{2}\left(1-\frac{1}{2N}\right)\Delta^{2}+\frac{\partial E_{2PPI}(m)}{\partial
    g^{2}}.
\end{equation}
This can be integrated using the ansatz
\begin{equation}\label{5}
    E=E_{2PPI}+cg^{2}\Delta^{2}.
\end{equation}
It remains to determine the unknown constant $c$.
Diagrammatically, it is easy to show that one has the following
gap equation.
\begin{equation}\label{6}
    \frac{\partial E_{2PPI}}{\partial m}=\Delta.
\end{equation}
Combination of the above formulae finally gives
\begin{equation}\label{7}
E=E_{2PPI}+\frac{1}{2}g^{2}\left(1-\frac{1}{2N}\right)\Delta^{2}.
\end{equation}
An important point is the renormalizability of the $2PPI$
expansion. Two possible problems could be mass renormalization and
vacuum energy renormalization, since originally there was no
external mass scale present. The proof is quite technical, but all
formulae remain correct and are finite when the conventional
counterterms of the Gross-Neveu model are included
\cite{Dudal:2002zn}. Essentially, the proof is based on coupling
constant renormalization and the separation of $2PPI$ and $2PPR$
contributions. \\\\It can be shown that
\begin{equation}\label{8}
    \frac{\partial E_{2PPI}}{\partial
    \Delta}=m\Leftrightarrow\frac{\partial E}{\partial m}=0.
\end{equation}
However, this does not mean that $E(\Delta)$ has the meaning of an
effective potential, since $E(\Delta)$ is meaningless if the gap
equation (\ref{8}) is not fulfilled.\\\\In Table 1, we list the
numerical deviations in terms of percentage between our optimized
2-loop results\cite{Dudal:2002zn} for the mass gap $M$ and the
square of minus the vacuum energy $\sqrt{-E}$ and the exact known
values. We conclude that the $2PPI$ results are in relative good
agreement with the exact values and converge to the exact
$N\rightarrow\infty$ limit.
\begin{table}[h]
\tbl{Numerical results for $M$ and $\sqrt{-E}$\vspace*{1pt}}
{\footnotesize
\begin{tabular}{|c|c|c|}
\hline
  $N$ & deviation $M$ (\%) & deviation $\sqrt{-E}$ (\%) \\
\hline
  2 & ? & ? \\
3 & -4.5 & 47.7 \\
4 & -6.5 & 27.9 \\
5 & -6.1 & 19.9 \\
10 & -3.5 & 8.4 \\
$\infty$ & 0 & 0\\
\hline
\end{tabular}\label{tab1} }
\vspace*{-13pt}
\end{table}
\section{$SU(N)$ Yang-Mills theory in the Landau gauge}
Next, we consider the Euclidean Yang-Mills action in the Landau
gauge where $A_{\mu}^a$ denotes the gauge field. Repeating the
analysis of section 2 leads to\cite{Dudal:2003vv}
\begin{eqnarray}
    \Delta&=&\left\langle
    A_{\mu}^{a}A_{\mu}^{a}\right\rangle,\nonumber\\
    m^{2}&=&g^{2}\frac{N}{N^{2}-1}\frac{d-1}{d}\Delta,\nonumber\\
  E &=& E_{2PPI}-\frac{g^{2}}{4}\frac{N}{N^{2}-1}\frac{d-1}{d}\Delta^{2},\nonumber\\
  \frac{\partial E_{2PPI}}{\partial m^{2}} &=&
  \frac{\Delta}{2}\Leftrightarrow\frac{\partial E}{\partial
  m^2}=0.
\end{eqnarray}
After some manipulation, the 2-loop results became
\begin{equation}
\frac{g^{2}N}{16\pi^{2}} \approx0.131\hspace{1cm} \sqrt{\Delta}
\approx 536 \textrm{MeV}\hspace{1cm} E\approx
-0.002\textrm{GeV}^{4}.
\end{equation}
We notice that the relevant expansion parameter, $\frac{g^2
N}{16\pi^2}$, is relatively small. As such, our results should be
qualitatively trustworthy. \\\\As a comparison, the value found by
Boucaud et al from lattice simulations and an OPE treatment was
$\left\langle A^2\right\rangle_{\mu=10\textrm{GeV}}\approx
1.64\textrm{GeV}$. To find a value for the gluon mass $m_{g}$
itself (as the pole of the gluon propagator) within the $2PPI$
framework, the diagrams relevant for mass renormalization of $m$
should be calculated.

\section*{Acknowledgments}
D.~D. would like to thank the organizers of this conference. This
work was supported in part by {\sc Pparc} through a research
studentship. (R.E.B.)

\end{document}